\documentclass{webofc} 
\usepackage[varg]{txfonts}   
\usepackage{hyperref} 
\usepackage{url} 
\hypersetup{colorlinks=true,citecolor=blue,urlcolor=blue,linkcolor=blue} 
\woctitle{MESON2026}

\def\lamb#1#2{$^{#1}_{\Lambda}${#2}}

\begin{document} 

\title{Comments on $\bf{\eta'}$ nuclei and on $B_{\Lambda}$(\lamb{3}{H})} 
\author{\firstname{Avraham} \lastname{Gal}\inst{1}\fnsep
\thanks{avragal@savion.huji.ac.il, invited talk at MESON2026, Krakow, 
June 2026.} 
\institute{Racah Institute of Physics, The Hebrew University, 
Jerusalem 9190401, Israel}} 

\abstract 
{I briefly discuss two apparent puzzles: (i) is the observation of 
$\eta'$(958) bound states in $^{11}$C at GSI \cite{GSI26} compatible 
with a nearly vanishing real part of $\eta'N$ scattering length \cite{FG25}? 
and (ii) is the $\Lambda$ binding energy of the hypertriton (\lamb{3}{H}), 
the lightest particle-stable hypernucleus, very small (63$^{+38}_{-31}$ keV, 
J-PARC E73 \cite{JPARC26}) or very large (523$\pm$76 keV, MAMI \cite{MAMI26})? 
Here, I suggest that MAMI's observed weak-decay pion momentum peak originates 
from \lamb{7}{He}, not from \lamb{3}{H} \cite{Gal26}.} 

\maketitle

\section{$\eta'$ nuclear bound states} 
\label{sec:eta'} 

\begin{figure}[!h] 
\centering 
\includegraphics[width=0.6\textwidth]{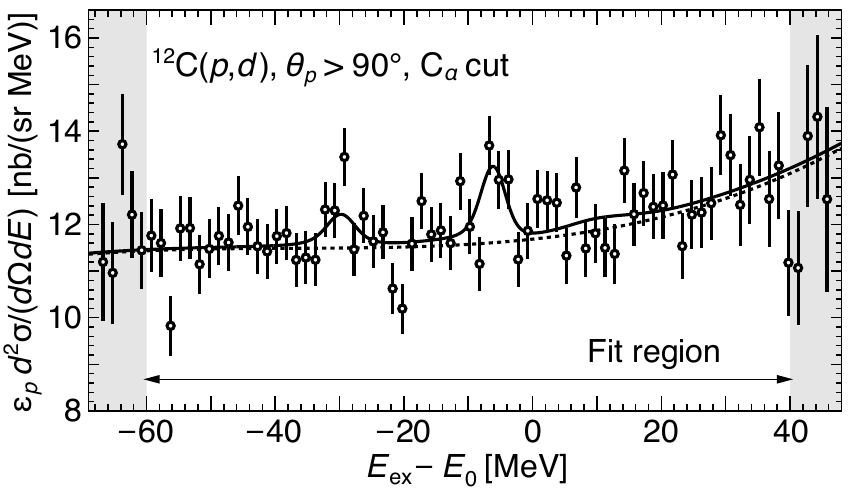} 
\caption{Excitation energy spectrum of the $^{12}$C(p,d) reaction near 
the $\eta'$ threshold $E_0$ measured recently at GSI, suggesting two 
$\eta'$-$^{11}$C states at 30 MeV and 6 MeV below threshold; figure adapted 
from Ref.~\cite{GSI26}. For an up-to-date overview of searches for $\eta'$ 
meson bound states in nuclei, see Metag's concluding talk here (MESON2026).} 
\label{fig:GSI26} 
\end{figure} 

There is considerable interest at present in establishing meson-nuclear 
quasibound states, particularly for ${\bar K}$-$\eta$-$\eta'$ pseudoscalar 
mesons and $\omega$-$\phi$ vector mesons, all with masses lower than or about
1~GeV~\cite{Metag17}. For quite some time the only meson-nuclear quasibound 
state established experimentally was ${\bar K}NN$, likely a ($J^P=0^-$, 
$I=\frac{1}{2}$) state, bound by a few tens of MeV but with a very large width 
exceeding 50~MeV~\cite{Yamaga20,Kobayakawa26}. In 2026, the observation of two 
relatively narrow deeply bound $\eta'$-$^{11}$C states in the $^{12}$C$(p,d)$ 
reaction at GSI was reported (see Fig.~\ref{fig:GSI26}), consistently with the 
in-medium $\eta'$ mass spectrum reported subsequently by SPring-8~\cite{SP26}.

\subsection{$\eta'$-nucleus optical potential} 
\label{subsec:Vopt} 

Bound states of $\eta'$ in nuclei were considered recently by Friedman and 
Gal~\cite{FG25}, using a density-dependent $\eta'$-nuclear optical potential 
\begin{equation}
V_{\rm opt}^{\eta'}(\rho) =
-\frac{4\pi}{2\mu_{\eta'}}f^{(2)}_A\,b_0^A(\rho)\,\rho,\,\,\,\,\,\, 
b_0^A(\rho) = \frac{{\rm Re}\,b_0}{1+(3k_F/2\pi)f^{(2)}_A{\rm Re}\,b_0} +
{\rm i}\,{\rm Im}\,b_0. 
\label{eq:V}
\end{equation}
Here, $b_0^A(\rho)$ is a density-dependent $\eta' N$ c.m. scattering amplitude 
in units of fm ($\hbar=c=1$), the nuclear density $\rho$ is normalized to the 
nuclear mass number $A$, $\rho_0=0.17$~fm$^{-3}$ is nuclear-matter density, 
$k_F=(3{\pi}^2\rho/2)^{1/3}$ is the Fermi momentum associated with the density 
$\rho$, $\mu_{\eta'}$ stands for the $\eta'$-nucleus reduced mass and 
$f^{(2)}_A$ is a kinematical factor transforming $b_0^A(\rho)$ from the 
$\eta'N$ c.m. system to the $\eta'$-nucleus c.m. system: 
$f^{(2)}_A=1+(1-1/A)(\mu_{\eta'}/m_N)$. 

The density dependence of $b_0^A(\rho)$ accounts for long-range Pauli 
correlations in $\eta' N$ in-medium multiple scatterings, starting at 
$\rho^{4/3}$ in a nuclear-density expansion. Note that Im$\,b_0$ in 
Eq.~(\ref{eq:V}) is not affected by Pauli correlations since all $\eta'N$ 
two-body decays proceed with c.m. momentum larger than $k_F$. In the 
low-density limit $b_0^A(\rho\to 0)\to b_0$ and $b_0$ is identified with the 
$\eta' N$ c.m. scattering length $a_{\eta' N}$. 

The $\eta'N$ scattering length $a_{\eta'N}$ is estimated from two 
near-threshold $\eta'$ production reactions on a free proton. 
Thus, COSY's $\sigma_{\rm tot}(pp\to pp\eta')$~\cite{COSY14} is fitted by  
\begin{equation}
{\rm Re}\,a_{\eta'p}=(0.00\pm 0.43)~{\rm fm} \,\,\,\,\,\,
{\rm Im}\,a_{\eta'p}=0.37^{+0.02+0.38}_{-0.11-0.05}~{\rm fm},  
\label{eq:cosy}
\end{equation} 
suggesting that Re$\,a_{\eta'p}$ is small w.r.t. Im$\,a_{\eta'p}$, while 
$\eta'$ photoproduction, $\gamma p\to\eta'p$~\cite{Anisovich18}, results in 
\begin{equation}
|a_{\eta'p}|=(0.403\pm 0.020\pm 0.060)\,\,\rightarrow \,\,
(0.356\pm 0.012)\,{\rm fm}, 
\label{eq:gamma} 
\end{equation} 
provided there is no narrow $s$-wave resonance near threshold (1896 MeV). 
Note that the $N^\ast$(1895) ($I=\frac{1}{2}, J^{\pi}={\frac{1}{2}}^-$) 
is quite broad ($\Gamma\approx\,$120 MeV~\cite{PDG26}).  

Optical potentials of the form Eq.~(\ref{eq:V}) where $b_0$ is identified with 
the $\eta'N$ scattering length $a_{\eta' N}$, assuming values consistent with 
Eqs.~(\ref{eq:cosy},\ref{eq:gamma}), fall short of binding $\eta'$ in 
$^{11}$C~\cite{FG25}. The corresponding $\eta'$ nuclear potential depth 
$D_{\eta'}$ is roughly given by 
\begin{equation} 
D_{\eta'}\equiv V_{\rm opt}^{\eta'}(\rho_0;A\to\infty)
\approx (-17\pm 6) -{\rm i}\,(27\pm 6)\,{\rm MeV} 
\label{eq:D} 
\end{equation}  
where, by Eq.~(\ref{eq:gamma}), the real and imaginary parts are highly 
correlated. 

We note that a situation where a meson-nucleon scattering length is dominantly 
imaginary is not often encountered in studies of mesic atoms and nuclei. 
For example, for $\eta$ mesons, various meson-baryon coupled-channel models 
of the broad ($I=\frac{1}{2}, J^{\pi}={\frac{1}{2}}^-$) $N^{\ast}$(1535) 
resonance, which is peaked about 50~MeV above the $\eta N$ threshold, all have 
Re$\,a_{\eta N}\gtrsim{\rm Im}\,a_{\eta N}$~\cite{Arndt05}. 

Regarding the sign of Re$\,a_{\eta'p}$, the (10-40)\% branching ratio
observed for $N^{\ast}(1895)\to N\eta'$ decay~\cite{PDG26} suggests that
$N^{\ast}$(1895), projected onto the $\eta' N$ channel, is a resonance rather 
than a bound state, implying in our phase convention that Re$\,a_{\eta'p}$ is 
positive.

\subsection{Implications of $^{12}$C(p,d) $\eta'$ bound-state GSI signals} 
\label{subsec:GSI} 

To establish contact with the recently reported $^{12}$C(p,d)$\eta'$ bound 
state signals that require $\eta'$ nuclear potential depth values~\cite{GSI26} 
Re~$D_{\eta'}\approx\,-61$~MeV and |Im~$D_{\eta'}$|$\,\leq\,$10~MeV we explore 
how large Re~$b_0$ should be in order to reproduce these signals assuming 
Im~$b_0$=0. Results of this exercise are listed in Table~\ref{tab:eta'}. 
\newpage 

\begin{table}[!t] 
\centering
\caption{$1s_{\eta'}$ and $1p_{\eta'}$ binding energies in $^{11}$C, 
calculated using $V_{\rm opt}^{\eta'}(\rho)$, Eq.~(\ref{eq:V}), 
for several values \newline of a real $\eta'N$ effective scattering length 
$b_0$, 
are listed together with $\eta'$ nuclear depth values $D_{\eta'}\equiv 
V_{\rm opt}^{\eta'}(\rho_0;A\to\infty)$, Eq.~(\ref{eq:D}). 
I thank Eli Friedman for providing me with this Table~\cite{FG26}.} 
\begin{tabular}{cccc} 
\hline
$b_0$ (fm) & $B_{1s}$ (MeV) & $B_{1p}$ (MeV) & $D_{\eta'}$ (MeV) \\
\hline
4.70 & 29.30 & 9.74 & $-$57.53 \\ 
2.65 & 24.15 & 6.01 & $-$51.92 \\ 
0.53 &  6.02 & ---  & $-$27.42 \\ 
\hline 
\end{tabular} 
\label{tab:eta'} 
\end{table} 

Values of $b_0$ that reproduce $B_{1s}\approx 30$~MeV or $B_{1p}\approx 6$~MeV 
are listed in the first two rows of Table~\ref{tab:eta'}. No single value 
of $b_0$ reproduces both, but quite interestingly the two rather different 
large values of $b_0$ give rise to $\eta'$ nuclear potential depths which 
are compatible with values $D_{\eta'}=-57.5^{-5.7}_{+27.8}$~MeV suggested by 
SPring-8~\cite{SP26}. Finally, the last row shows a fairly large reduction of 
$b_0$ and $|D_{\eta'}|$ values if only one $\eta'$ bound state exists with 
$B_{1s}\approx\,$6~MeV. Yet, the resulting value of $D_{\eta'}$ does not fall 
outside of 1$\sigma$ from the SPring-8 value.

\section{Questioning MAMI's extraction of $B_{\Lambda}(_{\Lambda}^3{\rm H})$ 
in $^7$Li$(e,e'K^+)$} 
\label{sec:L3H} 

\begin{figure}[!h] 
\centering 
\includegraphics[width=0.8\textwidth]{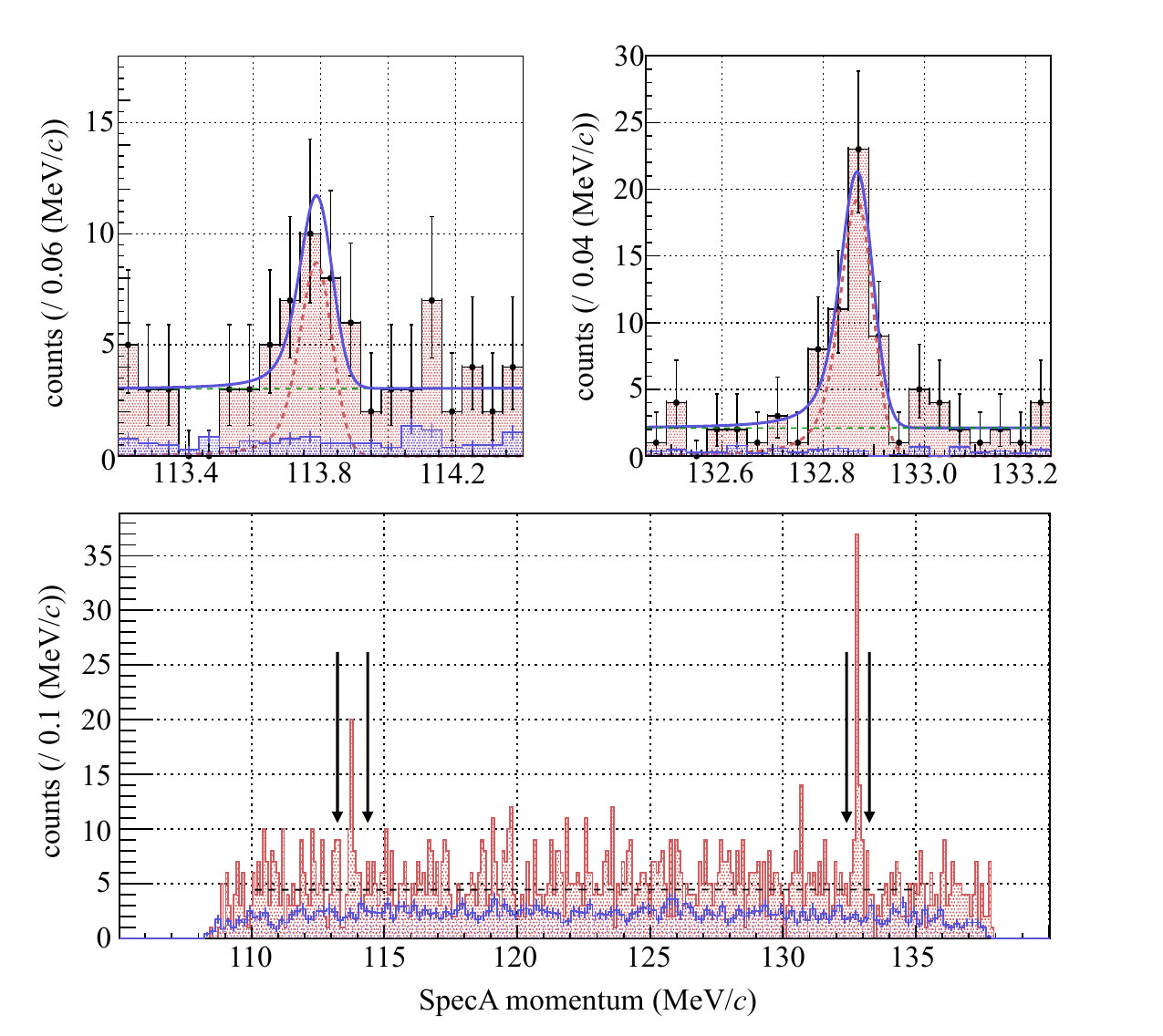} 
\caption{Pion weak-decay spectrum in $^7$Li$(e,e'K^+)$ from 
MAMI~\cite{MAMI26} with $p_{\pi^-}$ peaks attributed to 
$_{\Lambda}^3$H (left) and $_{\Lambda}^4$H (right).} 
\label{fig:MAMI} 
\end{figure} 

\begin{figure}[!h] 
\centering 
\includegraphics[width=0.7\textwidth]{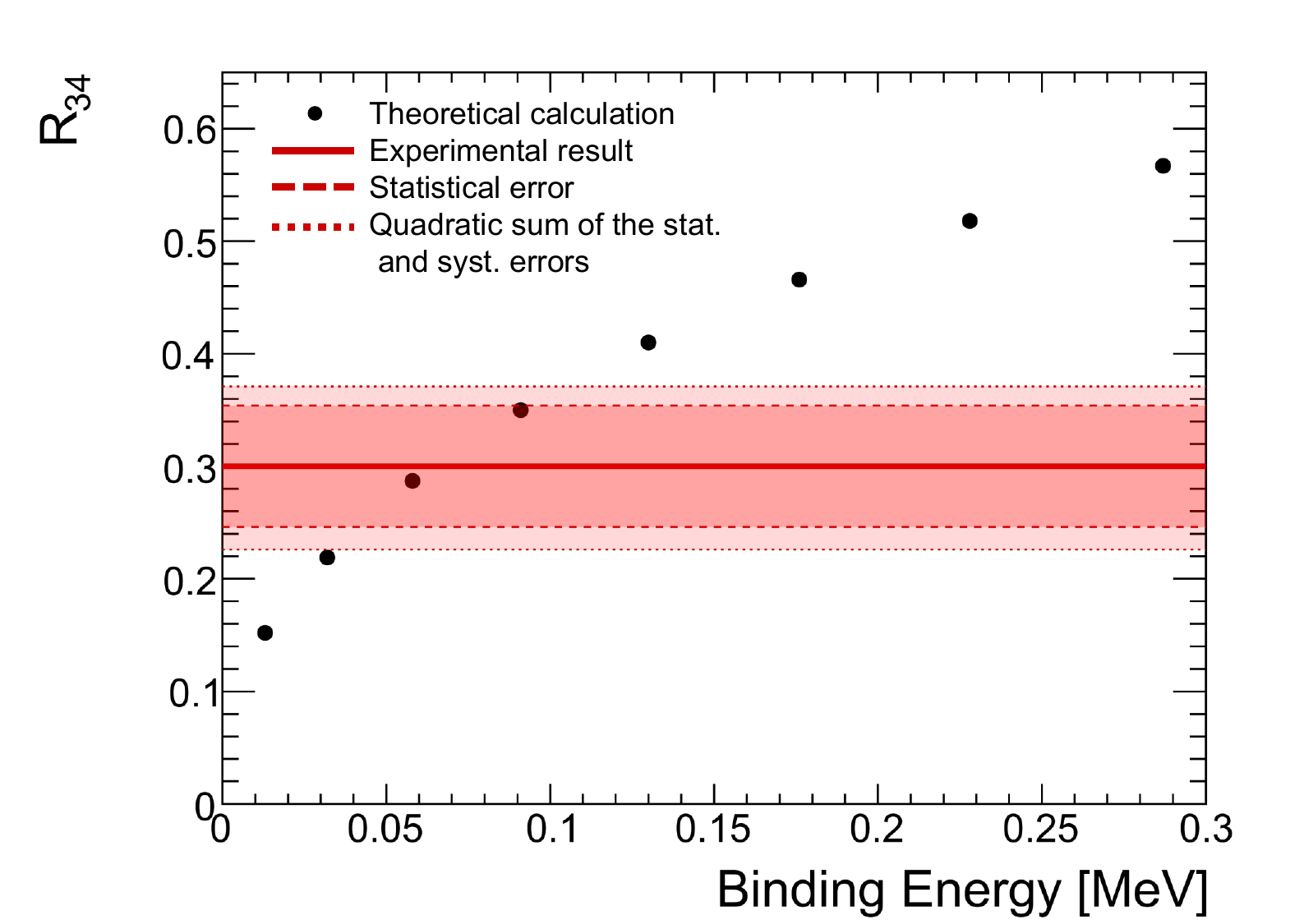} 
\caption{$^{3,4}_{\Lambda}$H production cross section ratio R$_{34}$ 
(red line) from J-PARC E73 $^{3,4}$He$(K^-,\pi^0)^{3,4}_{\Lambda}$H 
reactions \cite{JPARC26}. Black points are from calculations by Harada 
and Hirabayashi~\cite{HH26}.} 
\label{fig:JPARC} 
\end{figure}

A recent report by the A1 Collaboration at MAMI~\cite{MAMI26} on ${^7{\rm Li}}
(e,e'K^+)$ electroproduction runs assigns a sharp pion-momentum line at 
$p_{\pi^-}\approx 113.8\pm 0.1$ MeV/c to \lamb{3}{H}$\,\to\pi^-+{^3{\rm He}}$ 
weak decay, see Fig.~\ref{fig:MAMI}, resulting in $B_{\Lambda}$(\lamb{3}{H})$
=0.523\pm 0.013\pm 0.075$ MeV. This value disagrees sharply with the J-PARC 
E73~\cite{JPARC26} value $B_{\Lambda}$(\lamb{3}{H})$=0.063\pm 0.026\pm 0.023$ 
MeV, derived by comparing the measured $^{3,4}_{\Lambda}$H production cross 
section ratio R$_{34}$ in the forward two-body strangeness exchange reactions 
$^{3,4}$He$(K^-,\pi^0)^{3,4}_{\Lambda}$H with solid calculations~\cite{HH26}, 
see Fig.~\ref{fig:JPARC}. While both J-PARC and MAMI agree on $B_{\Lambda}
$(\lamb{4}{H}), first observed by MAMI in ${^9{\rm Be}}(e,e'K^+)$ 
electroproduction runs~\cite{MAMI15}, this is not the case for $B_{\Lambda}
$(\lamb{3}{H}). \lamb{3}{H} is a unique choice of final hypernuclear species 
on J-PARC E73's $^3$He target, but it may not hold in doing $(e,e'K^+)$ on 
$^7$Li. The question raised by me recently~\cite{Gal26} is why not 
\lamb{7}{He} for a final species? 

Two possible \lamb{7}{He}(1/2$^+_{\rm g.s.})\to\pi^-+{^7}$Li weak decays are 
listed in Table~\ref{tab:p(pi)}. Input hypernuclear masses were evaluated 
using binding energies $B_{\Lambda}$(\lamb{7}{He}), including uncertainties, 
in two ways: first from JLab's most recent ${^7{\rm Li}}(e,e'K^+)$\lamb{7}{He} 
electroproduction experiment~\cite{Gogami16}, with $p_{\pi^-}$ values listed 
in the third column; and second, in the fourth column, by requiring 
$p_{\pi^-}$=113.79$\pm$0.11~MeV/c for one of the two $^7$Li final states, 
in agreement with MAMI's sharp line in Fig.~\ref{fig:MAMI}. This leads to 
$B_{\Lambda}$(\lamb{7}{He})=5.84$\pm$0.07~MeV, almost 2$\sigma$ away from 
JLab's 5.55$\pm$0.15~MeV. Such a value would require also a similarly larger 
value of $B_{\Lambda}$(\lamb{7}{Li}) than the widely accepted emulsion value 
5.58$\pm$0.05~MeV, as discussed in Ref.~\cite{Gal26}. 

\begin{table}[!h]
\centering
\caption{Pion momentum values $p_{\pi^-}$ expected in $\pi^-$ weak decay 
of \lamb{7}{He}(1/2$^+_{\rm g.s.}$) to the two lowest $^7$Li bound states, 
(3/2$^-_{\rm g.s.}$) and (1/2$^-_{\rm exc.}$) at excitation energy 
$E_{\rm exc.}$=0.478~MeV, for two choices of $B_{\Lambda}$(\lamb{7}{He}): 
(i) 5.55$\pm$0.15~MeV~\cite{Gogami16} and (ii) 5.84$\pm$0.07~MeV; 
see text.}
\begin{tabular}{cccc}
\hline
\lamb{7}{He}: $J_i^{\pi}$ & $^7$Li: $J_f^{\pi}$ & $p^{(1)}_{\pi^-}$ (MeV/c) & 
$p^{(2)}_{\pi^-}$ (MeV/c) \\
\hline
(1/2$^+_{\rm g.s.}$) & (3/2$^-_{\rm g.s.}$) & 114.98$\pm$0.24 &
114.53$\pm$0.11 \\
(1/2$^+_{\rm g.s.}$) & (1/2$^-_{\rm exc.}$) & 114.24$\pm$0.24 &
113.79$\pm$0.11 \\
\hline 
\end{tabular} 
\label{tab:p(pi)} 
\end{table} 

We note that MAMI's measured pion spectra show no signal between $p_{\pi^-}$ 
values 114 to 115 MeV/c~\cite{MAMI26_com} where, see Table~\ref{tab:p(pi)}, 
the \lamb{7}{He}(1/2$^+_{\rm g.s.})\to\pi^-+{^7}$Li(3/2$^-_{\rm g.s.}$) weak 
decay is expected. Its rate is about twice stronger than the $\pi^-$ decay 
rate to $^7$Li(1/2$^-_{\rm exc.}$) at $E_x$=0.478~MeV, as follows from a 
similar case of $\pi^-$ weak decay rates of \lamb{7}{Li}(1/2$^+_{\rm g.s.}$) 
to $^7$Be(3/2$^-_{\rm g.s.}$) and $^7$Be(1/2$^-_{\rm exc.}$) at $E_x$=0.429 
MeV~\cite{Gal09}. However, $^7$Li(1/2$^-_{\rm exc.}$) decays by $\gamma$ 
emission to $^7$Li(3/2$^-_{\rm g.s.}$) almost instantaneously on a scale of 
hypernuclear lifetimes, thereby partly blocking $^7$Li(3/2$^-_{\rm g.s.}$) 
$m$ states that potentially contribute to the direct 
\lamb{7}{He}(1/2$^+_{\rm g.s.}$) $\pi^-$ weak decay to 
$^7$Li(3/2$^-_{\rm g.s.}$). This apparently leads to suppression of the 
latter transition such that the resulting $\pi^-$ weak decay rate to 
$^7$Li(3/2$^-_{\rm g.s.}$) is as small as 1/6-th of the weak decay rate to 
$^7$Li(1/2$^-_{\rm exc.}$)~\cite{Gal27}, consistently with the non-observation 
of the $\pi^-$ weak decay to $^7$Li(3/2$^-_{\rm g.s.}$)~\cite{MAMI26_com}. 

\begin{figure}[!h] 
\centering 
\includegraphics[width=0.7\textwidth]{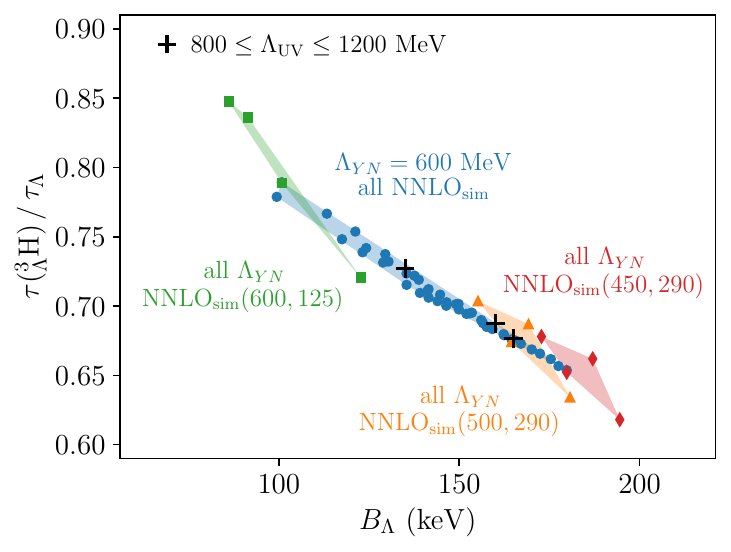} 
\caption{Correlation between \lamb{3}{H} lifetime $\tau$(\lamb{3}{H}) 
(in units of $\tau_{\Lambda}$=262$\pm$1~ps) and its $\Lambda$ binding energy 
$B_{\Lambda}$(\lamb{3}{H}), calculated using N$^2$LO $NN$ and LO $YN$  
$\chi$EFT interactions. I thank Daniel Gazda for providing me with this 
version of Fig.~7 in Ref.~\cite{Gazda24}.} 
\label{fig:TL3H}
\end{figure} 

Having argued that MAMI's sharp $p_{\pi^-}$ weak-decay line observed 
in ${^7{\rm Li}}(e,e'K^+)$ electroproduction (and assigned by MAMI 
to \lamb{3}{H}$\to \pi^- + ^3$He~\cite{MAMI26}) is more likely 
to arise from $\pi^-$ weak decay of \lamb{7}{He}~\cite{Gal26} 
removes the tension between a very large MAMI's $B_{\Lambda}$(\lamb{3}{H}) 
value of order 0.5~MeV and the fairly small J-PARC E73's value 
$B_{\Lambda}$(\lamb{3}{H})=0.063$\pm$0.026$\pm$0.023 MeV~\cite{JPARC26}. 
Theoretically, a small $B_{\Lambda}$(\lamb{3}{H}) value is correlated with 
\lamb{3}{H} lifetime $\tau$(\lamb{3}{H}) values close to the free-$\Lambda$ 
lifetime $\tau_{\Lambda}$, as shown in Fig.~\ref{fig:TL3H} for $\chi$EFT 
calculations using $NN$ and $YN$ interactions to order N$^2$LO and LO, 
respectively~\cite{Gazda24}. 
Experimentally, the Chart of Hypernuclides~\cite{KPH} suggested value 
for $\tau$(\lamb{3}{H}) is 238$^{+9}_{-8}$ ps (dominated by ALICE's value 
253$\pm$11$\pm$6~ps~\cite{ALICE23}) and is correlated theoretically by 
Fig.~\ref{fig:TL3H} with $B_{\Lambda}$(\lamb{3}{H})$\leq$100 keV which is 
clearly inconsistent with MAMI's $B_{\Lambda}$(\lamb{3}{H})$\sim$500 keV.

\section{Summary} 
\label{sec:sum} 

The present discussion of $\eta'$ nuclear bound states in the first half of 
the talk was triggered by somewhat surprising results of two very recent 
experimental reports, one from GSI~\cite{GSI26} and another one from 
SPring-8~\cite{SP26}, both suggesting a deeply attractive $\eta'$-nuclear 
potential of depth about 60~MeV. This conclusion requires that both GSI's 
bound-state signals are confirmed as genuine $\eta'$-$^{11}$C bound states. 
A deeply attractive $\eta'$-nuclear potential contrasts with the weakly 
attractive and quite absorptive $\eta'$-nuclear optical potential constrained 
by experimental values of $\eta'N$ scattering length $a_{\eta'N}$~\cite{FG25}. 
We then mentioned very briefly the potential role of the N$^\ast$(1895) 
resonance in making the $\eta'N$ scattering amplitude below threshold 
compatible with such bound state signals. 

In the second half of the talk I questioned MAMI's recent claim of 
a fairly large value of the $\Lambda$ binding energy in \lamb{3}{H}, 
$B_{\Lambda}$(\lamb{3}{H})$\approx$0.5~MeV~\cite{MAMI26}, based on 
assigning a sharp line at $p_{\pi^-}\approx\,$113.8~MeV/c observed in 
${^7{\rm Li}}(e,e'K^+)$ to the two-body weak decay \lamb{3}{H}$\to\pi^-+{^3}
$He. This sharp line could be attributed as well to the weak decay 
\lamb{7}{He}(1/2$^+_{\rm g.s.}$)$\to\pi^-+^7$Li(1/2$^-_{\rm exc.}$). 
The main difficulty with this interpretation is that it implies naively 
also a sharp $\pi^-$ weak decay line at $p_{\pi^-}\approx\,$114.5~MeV/c 
to $^7$Li(3/2$^-_{\rm g.s.}$) which is not observed. To resolve 
this difficulty I noted that the fast $M1$ $\gamma$ transition 
$^7$Li(1/2$^-_{\rm exc.}$)$\to^7$Li(3/2$^-_{\rm g.s.}$) apparently leads to 
suppression of the $p_{\pi^-}\approx\,$114.5~MeV/c $\pi^-$ weak decay line 
to the same final state. I then presented some theoretical arguments that 
advocate a rather small value of $B_{\Lambda}$(\lamb{3}{H}), about 0.1~MeV, 
supported experimentally by the J-PARC E73~\cite{JPARC26} reported small value 
of $B_{\Lambda}$(\lamb{3}{H})$\leq$0.1~MeV and by ALICE's precise lifetime 
determination of a $\tau$(\lamb{3}{H}) value close to $\tau_{\Lambda}$. Thus, 
I question the assignment of \lamb{3}{H} to the sharp $\pi^-$ weak-decay line 
observed in the ${^7{\rm Li}}(e,e'K^+)$ reaction spectrum reported recently by 
MAMI~\cite{MAMI26}.

\section*{Acknowledgments} 
\begin{acknowledgement} 
The work on $\eta'$ nuclei reviewed here was done in collaboration with Eli 
Friedman. Discussions with Volker Metag on meson-nuclear bound states are 
gratefully acknowledged. Thanks are also due to the organizers of MESON2026 
in Krakow for support and hospitality. This work started as part of a project 
funded by the EU Horizon 2020 Research \& Innovation Programme under grant 
agreement 824093. 
\end{acknowledgement}

\end{document}